\documentstyle[12pt,fleqn]{article}
\input epsf
\def\be{\begin{equation}}
\def\ee{\end{equation}}
\def\ba{\begin{eqnarray}}
\def\ea{\end{eqnarray}}
\def\la{\mathrel{\mathpalette\fun <}}
\def\ga{\mathrel{\mathpalette\fun >}}
\def\fun#1#2{\lower3.6pt\vbox{\baselineskip0pt\lineskip.9pt
        \ialign{$\mathsurround=0pt#1\hfill##\hfil$\crcr#2\crcr\sim\crcr}}}
\def\pho{{{\widetilde{\gamma}}}}
\def\r0{{{R^0}}}
\def\rpi{{{R_\pi}}}
\def\glu{{{\widetilde{g}}}}

\def\GeV{{{\mbox{GeV}}}}

\def\mb{{{\mbox{mb}}}}
\def\avg#1{{{{\langle #1 \rangle }}}}

\def\re#1{{[\ref{#1}]}}
\def\eqr#1{{Eq.\ (\ref{#1})}}
\def\mst{{{M_{S}}}}

\newcommand{\intcvt}{$\pho-\r0$ conversion }

\begin{document}
\begin{titlepage}
\null\vspace{-72pt}
\begin{flushright}
{\footnotesize
FERMILAB--Pub--96-097-A\\
RU-97-13\\
astro-ph/9703145\\
March 1997 \\
}
\end{flushright}
\renewcommand{\thefootnote}{\fnsymbol{footnote}}
\vspace{0.15in}
\baselineskip=24pt

\begin{center}
{\Large \bf On the relic abundance of light photinos}
\baselineskip=14pt
\vspace{0.75cm}

Daniel J.\ H.\ Chung\footnote{Electronic mail: 
		{\tt djchung@yukawa.uchicago.edu}}\\
{\em Department of Physics and Enrico Fermi Institute\\ 
The University of Chicago, Chicago, Illinois~~60637, and\\
NASA/Fermilab Astrophysics Center\\
Fermi National Accelerator Laboratory, Batavia, Illinois~~60510}\\
\vspace{0.4cm}
Glennys R.\ Farrar\footnote{Electronic mail: 
		{\tt farrar@physics.rutgers.edu}}\\
{\em Department of Physics and Astronomy\\
Rutgers University, Piscataway, New Jersey~~08855}\\
\vspace{0.4cm}
Edward W.\ Kolb\footnote{Electronic mail: 
	{\tt rocky@rigoletto.fnal.gov}}\\
{\em NASA/Fermilab Astrophysics Center\\
Fermi National Accelerator Laboratory, Batavia, Illinois~~60510, and\\
Department of Astronomy and Astrophysics and Enrico Fermi Institute\\
The University of Chicago, Chicago, Illinois~~60637}\\
\end{center}
\baselineskip=24pt

\begin{quote}
\hspace*{2em}
We solve the coupled Boltzmann equation for the system of light
photinos interacting with pions and $\r0$'s (the gluon-gluino bound
state) to determine the relic abundance of light photinos in
the light gaugino scenario.  Cosmology bounds the ratio $r$ of the
$\r0$ mass to the $\pho$ mass to be less than about $1.8$.
We also use a model Lagrangian embodying crossing symmetry between the
$\r0 \leftrightarrow \pho \pi \pi$ and $\r0 \pi \leftrightarrow \pho
\pi$ reactions to identify cosmologically favored regions of $\r0$
lifetime as a function of $\r0$ and $\pho$ masses.
\vspace*{8pt}

PACS number(s): 95.35.+d, 14.80.Ly, 98.80.Cq

\renewcommand{\thefootnote}{\arabic{footnote}}
\addtocounter{footnote}{-2}
\end{quote}
\end{titlepage}

\newpage

\baselineskip=24pt
\renewcommand{\baselinestretch}{1.5}
\footnotesep=14pt

\vspace{36pt}
\centerline{\bf I. INTRODUCTION}
\vspace{24pt}
In supersymmetric (SUSY) models without dimension-3
supersymmetry-breaking operators, gauginos are massless at the tree
level and obtain non-zero masses solely from radiative corrections
[\ref{bgmbm},\ref{FM},\ref{pierce_papa}].  This means that the gluino
is light and the lightest neutralino is nearly a pure photino.  Farrar
[\ref{EXPTS},\ref{PHENO},\ref{fPRL}] found that the light gluinos and
photinos arising from this scenario are consistent with the present
experimental constraints.\footnote{The recent ALPEH claim to exclude
light gluinos [\ref{aleph:lg}] assigns a $1 \sigma$ theoretical
systematic error based on varying the renormalization scale over a
small range. Taking a more generally accepted range of scale variation
and accounting for the large sensitivity to hadronization model, the
ALEPH systematic uncertainty is comparable to that of other
experiments and does not exclude light gluinos [\ref{fLaT}].}  Although
it was once generally believed that the light gaugino scenario
conflicted with the cosmological relic abundance constraints, Farrar
and Kolb \re{farkolb} showed that the previous constraint calculations
neglected the reaction channels which really control the relic
abundance.  Indeed, based on some simple estimates, they concluded
that a light photino (the relic stable particle in the present SUSY
scenario) might be a significant dark matter candidate.  However,
their estimates were based on the approximation that only a single
reaction dominates the relic abundance evolution and that the
abundance evolution stops exactly when the dominant reaction rate
becomes less than the Hubble expansion rate (the ``sudden''
approximation). 

In the present paper, we calculate the cosmological constraints for
this scenario of light gluinos and photinos more carefully by
integrating the Boltzmann equations for the relic abundance.  The three
most important reactions determining the photino abundance are $R^0
\pi^{\pm} \leftrightarrow \pi^{\pm} \pho$, $R^0 \rightarrow \pi^+
\pi^- \pho$, and  $ R^0 R^0 \rightarrow X$.  The first two are related
by crossing symmetry.  In the limit that left and right handed squark
masses are equal, such that charge conjugation is a good symmetry of
the theory, and ignoring the momentum dependence of the matrix
elements, this crossing relation indicates that both reactions may
play an important role instead of one reaction dominating over the
other.  We use the results of our model calculations to help identify
the cosmologically most promising values for phenomenologically
important parameters such as the $\r0$ lifetime, which can help in
laboratory searches. 

Let us now briefly introduce the relevant features of our SUSY
scenario.  Supersymmetric models with acceptable SUSY breaking
phenomenology are generically invariant under a global chiral symmetry
called $R$-invariance.  $R$-invariance is broken spontaneously by the
vacuum expectation values of the Higgs fields associated with
electroweak symmetry breaking, and by tree-level gaugino masses if
they are present.  $R$-parity is the possible discrete remnant of this
broken continuous symmetry.  Under $R$-parity, the gluino, photino,
and squarks are odd, while ordinary particles (e.g., gauge and Higgs
bosons and quarks) are even.  $R$-parity, which we shall assume is an
unbroken symmetry, ensures that the lightest $R$-odd particle is
stable and prevents unacceptably rapid proton decay.  Thus, in
calculating the relic density in SUSY, one first identifies the
lightest $R$-odd particle which usually is the lightest supersymmetric
particle (LSP).  Although the gluino may be the lightest particle in
our scenario, it cannot exist in isolation today because it is not a
color singlet. Bound to a gluon or a color-octet system of quarks
and/or anti-quarks, it forms a color singlet hadron.  The lightest of
these is expected to be a gluon-gluino bound state called $R^0$, whose
mass should be comparable to that of the lightest glueball
[\ref{EXPTS}, \ref{PHENO}].  Because this is most likely heavier than
the photino, it is the photino which acquires the role usually taken
on by the LSP even though it may be heavier than the gluino.\footnote{
In some SUSY-breaking models only the gluino is massless at tree
level, while other gauginos have large masses.  In this case the $\r0$
could be the LSP and relic $\r0$'s would be the SUSY dark matter
candidate.  The dark matter density can be approximated as in \re{KT},
accounting for only the $\r0 \r0$ self-annihilation.  This gives
$\Omega_\r0 h^2 \la 10^{-7}$.  That is, due to their strong
interactions, $\r0$'s stay in thermal equilibrium too long for their
abundance to freeze out at a non-negligible value. Thus such
SUSY-breaking scenarios do not provide a natural visible sector dark
matter candidate unless the gravitino has acceptable properties.}

Since freeze out occurs after the color confinement phase transition,
only gluinos bound in color singlet states are relevant to our
calculation.  Among the bound states containing a gluino
($R$-hadrons), the $\r0$ is expected to react most prominently with
the photino because other $R$-hadrons are significantly heavier and
thus Boltzmann suppressed at the relevant temperatures.  Furthermore,
most of the other $R$-odd states will contribute to the photino
abundance only after having decayed to an $\r0$ channel.  Thus, the
photino relic abundance will be determined primarily by the reactions
involving an $\r0$, a $\pho$, and non-SUSY particles.

\begin{table}
\footnotesize{\hspace*{1em} Table I: A list of SUSY mass
parameters and ranges used in the analysis.}
\begin{center}
\begin{tabular}{lccc}
Particle & Mass Notation & Min.(GeV) & Max.(GeV) \\
\hline \hline
photino($\pho$) & $m$ & 0.2 & 1.4 \\
$\r0(\glu g)$ & $M$ & 1 & 2 \\
squark & $M_S$ & 50 & 300 \\
\hline \hline
\end{tabular}
\end{center}
\end{table}

In our scenario, the photino abundance depends crucially on
interactions of hadrons after the confinement phase transition,
causing complications distinct from conventional scenarios where the
freeze out occurs above the confinement transition temperature.  In
particular, we are only able to make reasonable guesses for the
relevant reaction rates because of incalculable long-distance QCD
effects and our lack of direct experimental data for the reaction
rates of interest.  Fortunately, we are still able to make useful
predictions regarding the $\r0$ and $\pho$ masses and $\r0$ lifetime.

The relic abundance of photinos depends mainly upon the masses of
$\pho$ and $\r0$, the cross sections for $\r0 \pi \rightarrow
\pi \pho$ and $\r0 \r0 \rightarrow X$ (where $X$ denote any strongly
interacting light species of particles such as the pions), and the
decay rates for $\r0
\rightarrow \pho \pi \pi$ and 
$\r0 \rightarrow \pho \pi $.  The mass parameter space that will be
explored in this calculation is justified in Refs.\ [\ref{FM}] and
[\ref{PHENO}] and is similar to that discussed in Ref.\ \re{farkolb}.
The relevant mass parameters and their plausible ranges are shown in
Table I.  The gluino mass itself is unimportant, except insofar as it
influences the $\r0$ mass, $M$.  The relevant squark mass denoted
$\mst$ is a charge-weighted average of up- and down-squark masses.
See Ref.\ \re{SQUARKS} for squark mass limits in the light gluino
scenario.

In order to express some of the formulae showing numerical estimates
concisely, we also define the following dimensionless ratios:
\be
\mu_8  \equiv  \frac{m}{0.8\ \GeV}; \qquad
\mu_{S} \equiv \frac{\mst}{100\ \GeV}; \qquad r\equiv \frac{M}{m}. 
\label{mparams}
\ee
As pointed out in Ref.\ \re{farkolb}, the relic abundance is
particularly sensitive to the parameter $r$.  Note that because of the
ranges we adopt, given in Table I, the range of $r$ we explore is
constrained for a fixed value of $m$. 

In the next section, we discuss the Boltzmann equation and some
simplifying assumptions used to calculate the present photino
abundance (density).  In Section III, we briefly describe the
reactions that are included in the simplified Boltzmann equations.  The
results of the integration are presented and analyzed in Section IV.
In Section V we develop an effective Lagrangian description of the
interaction between $\r0$, $\pho$ and pions which embodies the
symmetries of the underlying theory as well as the crossing and
chiral-perturbation theory constraints.  Ignoring the possibility that
a nearby $\rpi$ resonance produces a strong momentum dependence, the
two dominant reactions controlling the $\pho$ abundance are determined
by a single parameter.  Using this approximation, we obtain an estimate
of the cosmologically favored lifetime range of the $\r0$ as a
function of its mass.  We summarize our results in Section VI.  In an
appendix, we analyze the possible resonance enhancement of the $\r0
\pi \rightarrow \pho \pi$ cross section using a Breit-Wigner model.

\vspace{36pt}
\centerline{\bf II. THE BOLTZMANN EQUATIONS}
\vspace{24pt}
The standard method of calculating the relic abundance is to integrate
a simplified form of the Boltzmann equations [\ref{GG},\ref{KT}].  We
now briefly remind the reader of the general formulation.  One can
write the Boltzmann equations\footnote{As usual, we have used the
assumption of molecular chaos to obtain a closed set of equations.}
for the evolution of the particle density $n_j$ as
\begin{eqnarray}
\frac{dn_j}{dt}+3H n_j & = & - \sum_i \frac{ \langle W_{j A_{ji}
\rightarrow B_{ji}} 
V^{n(A_{ji})} \rangle }{\prod_{\lambda \in B_{ji}} n_\lambda^{eq}}
\nonumber \\
& & \times \left(n_j \prod_{k \in A_{ji}} n_k \prod_{\lambda \in B_{ji}} 
n_\lambda^{eq}-n_j^{eq} \prod_{k \in A_{ji}} n_k^{eq} \prod_{\lambda \in
B_{ji}} n_\lambda\right)
\label{eq:boltzmaster}
\end{eqnarray}
where $H$ is the usual Hubble expansion rate, $A_{ji}$ and $B_{ji}$
are sets of particle species relevant to the evolution of
species $j$, and the summation is over all the reactions of the form $j
A_{ji} \rightarrow B_{ji}$.  We have defined the thermal averaged
transition rate as
\begin{eqnarray}
\lefteqn{\langle W_{j A_{ji} \rightarrow B_{ji}} V^{n(A_{ji})} \rangle
 \equiv \frac{{\displaystyle \int \left[ dp \right] (2 \pi)^4
 \delta^{(4)}\left( \sum_k p_k \right) |T_{j A_{ji} \rightarrow B_{ji}}|^2
 \exp(- \sum_{\lambda \in j, A_{ji}} E_\lambda/T)}}{ \prod_{p
 \in j,A_{ji}} (n^{eq}_p/g_p)}} \\ 
\lefteqn{ \left[dp \right] \equiv
 \prod_{k \in j,A_{ji},B_{ji}} \frac{d^3 p_k}{(2 \pi)^3 2 E_k}} \\
\lefteqn{n^{eq}_x \equiv g_x \int \frac{d^3p_x}{(2 \pi)^3} \exp(-E_x/T)},
\label{eq:equilibn} 
\end{eqnarray}
where $n_x^{eq}$ is the equilibrium density,\footnote{As usual, the
particle described by this equilibrium density is assumed to have mass
much greater than the temperature.} $g_x$ counts the spin
multiplicity, and $|T_{j A_{ji} \rightarrow B_{ji}}|^2$ represents the
spin averaged transition amplitude squared.\footnote{The $W$ symbol
actually represents the number of transitions per unit time when all
the initial state reactants have equal unit flux.  The $V$ symbol
represents the characteristic spatial volume of interaction and the
power $n$ is the number of initial state particles minus one.} In the
case of one initial state particle, $\langle W_{j A_{ji} \rightarrow
B_{ji}}V^{n(A_{ji})} \rangle$ evaluates to a decay rate, whereas in
the case when there are two initial state particles, $\langle W_{j
A_{ji} \rightarrow B_{ji}}V^{n(A_{ji})} \rangle $ evaluates to the
familiar $\langle\sigma v\rangle$ of a scattering reaction.  For
example, in the case of a photino density evolution determined only by
the reaction $ \pho \pho \leftrightarrow X$, the density labels become
$j=\pho$, $A_{ji}= A_{\pho \pho}= \{ \pho \}$, and $B_{j i}=B_{\pho
\pho}=\{ X \}$; the summation in $i$ reduces to a sum over one element (the
annihilation channel); and the transition rate per unit fluxes becomes
$\langle W_{\pho \pho \rightarrow X} V \rangle = \avg{v \sigma(\pho
\pho \rightarrow X)}$.  With the usual assumption that the final
products $X$ are in equilibrium, \eqr{eq:equilibn} reduces to the
familiar equation (see, for example, pg.\ 120 of Ref.\ \re{KT})
\be
\frac{dn_{\pho}}{dt}+3 H n_{\pho}=-\avg{v \sigma(\pho \pho \rightarrow
X)} (n_{\pho}^2-n^{eq \, 2}_{\pho}).
\ee
Note that \eqr{eq:boltzmaster} assumes that the fluid is rare enough to
disregard degenerate pressure effects and assumes that time reversal
is a good symmetry.  More specifically, time reversal symmetry is
encoded in the following identity used in obtaining \eqr{eq:boltzmaster}:
\be
\langle W_{j A_{ji} \rightarrow B_{ji}}V^{n(A_{ji})} \rangle= \langle
W_{B_{ji} \rightarrow j A_{ji}}V^{n(B_{ji})-1} \rangle
\frac{\prod_{\lambda \in B_{ji}}
n_\lambda^{eq}}{n_j^{eq} \prod_{\lambda 
\in A_{ji}} n_\lambda^{eq}} 
\label{eq:timerev}
\ee

Before we can utilize \eqr{eq:boltzmaster} to determine the relic
abundance of the photinos, we need to specify our model of $H$ and the
reactions that are involved.  Because the universe is radiation
dominated for the temperatures of interest, the equation of state is
taken to be $3\times(\mbox{pressure})=(\mbox{energy density})$ and any
possible spatial curvature is neglected.  We also use equilibrium
statistics with the number of relativistic degrees of freedom set to
$g_*=10.75$.  The resulting equation for the Hubble expansion rate as
a function of temperature is $H=\sqrt{8 \pi^3 g_*/90}(T^2/m_{pl})$,
where $m_{pl}$ is the Planck mass.  The reactions that can enter the
Boltzmann equations include $\pho \r0
\leftrightarrow X$, $\pho \pho \leftrightarrow X$, $\r0 \r0
\leftrightarrow X$, $\pho \pi \leftrightarrow \r0$, $\pho \pi \pi
\leftrightarrow \r0$,  and $\pho \pi \leftrightarrow \r0 \pi$ ($X$'s
denote any allowed light products that interact strongly or
electromagnetically).  

In general, \eqr{eq:boltzmaster} generates a set of coupled nonlinear
differential equations which can be solved numerically.  However,
instead of considering all the particle densities as unknowns, we can
simplify the situation with the good approximation that the particle
densities whose equilibrating chemical reaction rate is large compared
to the Hubble expansion rate follow equilibrium densities of the form
\eqr{eq:equilibn}.  This, in fact, is the justification for our
Boltzmann evolution's initial condition which is to start all species
at equilibrium densities given by \eqr{eq:equilibn}.  With this
expectation, we replace the $X$ and the $\pi$ densities in the
Boltzmann equation with the equilibrium densities.  We are then left
to consider only the $\r0$ and the $\pho$ densities as functions that
require solutions.

To understand which reactions will be most important in our system, we
first recast \eqr{eq:boltzmaster} into the dimensionless form
\be
\frac{x}{Y_j}\frac{d}{dx}Y_j= - \sum_i 
\frac{\langle W_{jA_{ji} \rightarrow
B_{ji}} V^{n(A_{ji})}\rangle\prod_{\lambda \in A_{ji}}
n_\lambda^{eq}}{H(x)} \left(\frac{ \prod_{\lambda \in A_{ji}}
Y_\lambda}{ \prod_{\lambda
\in A_{ji}} Y_\lambda^{eq}} -  
\frac{Y_j^{eq} \prod_{k \in B_{ji}} Y_k}{Y_j \prod_{k \in B_{ji}}
Y_k^{eq}} \right) 
\label{eq:boltz2}
\ee where $Y_r=n_r/s$, $s$ is the entropy per comoving volume given by
$s  \approx (2 \pi^2/45) g_* m^3/x^3$ (entropy conservation is
assumed), and $x=m/T$.  Note that we can interpret the numerator above
$H(x)$ to be the reaction rate per unit density of $j$'s.  For the
purpose of illustration, suppose two reactions named $a$ and $b$ are
governing the evolution of $j$ particles and the reaction rates
corresponding to them are labeled $R_a$ and $R_b$.  The evolution
equation in the form of \eqr{eq:boltz2} then becomes 
\be
\frac{x}{Y_j}\frac{d Y_j}{dx}=-\frac{R_a(x)}{H(x)}(\mbox{ratios
a})-\frac{R_b(x)}{H(x)}(\mbox{ratios b}) 
\ee 
where the ``ratios''
refer to the terms consisting of density ratios.  Suppose further that
we are at a time when $R_a(x)/H(x) \ll 1$ while $R_b(x)/H(x) \gg 1$.
Then as long as the ``ratios a'' and ``ratios b'' are comparable in
value, reaction $a$ can be neglected during this period of evolution.
Furthermore, if the final products of reaction $b$ are in equilibrium,
the $j$ particle density will follow the equilibrium density as long
as reaction $b$ dominates.  With such reasoning, Ref.\ \re{farkolb}
argues that $\pho \r0\leftrightarrow X$ and $\pho \pho \leftrightarrow
X$ reactions play a negligible role compared to $\r0 \r0
\leftrightarrow X$, $\pho \pi \leftrightarrow \r0 \pi$, and $\pho \pi
\leftrightarrow \r0$ in keeping the $\r0$ and the $\pho$ densities in
equilibrium near the time of $\pho$ freeze out.  In our present work,
we shall neglect only the weakest of the relevant reactions, $\pho \r0
\leftrightarrow X$.\footnote{We have checked numerically that this
reaction plays a negligible role.}

The Boltzmann equations relevant to calculating the $\pho$ abundance
thus reduce to a pair of coupled differential equations containing
terms corresponding to the set of reactions $\pho\pi \pi
\leftrightarrow \r0$, $\pho\pi \leftrightarrow
\r0$, $\pho \pi \leftrightarrow \r0 \pi$, $\r0 \r0 \leftrightarrow X$,
$\pho \pho \leftrightarrow X$: 
\begin{eqnarray}
\frac{x}{Y_\pho}\frac{dY_\pho}{dx} & = &
-\frac{R_{tot}}{H}\left( 1- \frac{Y_\pho^{eq}
Y_\r0}{Y_\pho Y_\r0^{eq}} \right) - \frac{2 \avg{W_{\pho \pho
\rightarrow X} V} n_\pho^{eq}}{H} \left(\frac{Y_\pho^{eq}}{Y_\pho}
\right) \left(
\frac{Y_\pho^2}{Y_\pho^{eq 2}}-1 \right)
\label{eq:boltz3} 
\\
\frac{x}{Y_\r0}\frac{dY_\r0}{dx}  & = & -\frac{R_{tot}}{H} \left(
\frac{Y_\pho^{eq}}{Y_\r0^{eq}} \right) \left(1- \frac{Y_\pho
Y_\r0^{eq}}{Y_\pho^{eq} Y_\r0} \right) \nonumber 
\\
& &- \frac{2
\avg{W_{\r0 \r0 \rightarrow X} V} n_\r0^{eq} }{ H(x)}
\left(\frac{Y_\r0^{eq}}{ Y_\r0} \right) \left(
\frac{Y_\r0^2}{Y_\r0^{eq 2}} - 1 \right)
\\
R_{tot} & \equiv & (\avg{W_{\pho \pi \rightarrow \r0} V}n^{eq}_{\pi} +
\avg{W_{\pho \pi \pi \rightarrow \r0} V^2}n^{eq}_{\pi}
n^{eq}_{\pi} + N \avg{W_{\pho \pi \rightarrow \pi \r0} V}
n^{eq}_{\pi}).
\label{eq:R_tot}
\end{eqnarray}
The factor of $N$ comes from summing over the isospins of $\pi$.  In
the next section, we argue that only $\pi^\pm$ should be included in
$\pho \pi \rightarrow \pi \r0$, resulting in $N=2$.\footnote{The
choice $N=3$ was implicit in the treatment in Ref.\ \re{farkolb}.}

Before we move on to discuss the transition rates, let us clarify the
term ``freeze out time'' used in this paper, particularly in Section
IV.  In agreement with what will be revealed in the next section,
suppose that the self-annihilation term in \eqr{eq:boltz3} can be
neglected compared to the term associated with $R_{tot}$.  When
$R_{tot}/H$ becomes much less than unity and continues decreasing
sufficiently fast to keep the right hand side of
\eqr{eq:boltz3} much less than unity despite the increases in the
magnitude of the term multiplying $R_{tot}/H$, the fractional change
in $Y_\pho$ becomes negligible.  This is then a sufficient condition
for the number of $\pho$'s becoming approximately constant (freezing
out).  We shall use the term freeze out time to refer to the
approximate time at which $R_{tot}/H$ becomes much less than unity.

\vspace{36pt}
\centerline{\bf III. THE TRANSITION RATES}
\vspace{24pt}

Transition amplitudes for $\r0$, $\pho$, and pions depend on hadronic
matrix elements of four-fermion effective operators of the form
$\tilde{g} \pho \bar{q} q$, obtained by integrating out the squark
degree of freedom.  Since only a small number of fundamental
short-distance operators underly all the transition amplitudes of
interest, crossing symmetry can be used to relate transition
amplitudes for some of the reactions.  Due to the possibly strong
momentum dependence of the amplitudes, however, this proves to be of
limited utility.  This is discussed in Section V.

The particles $\r0$ and $\pho$ are charge conjugation even and odd,
respectively [\ref{fPRL}].  Thus, if charge conjugation were a good
symmetry of the interaction, pions coupling an $\r0$ to a $\pho$ would
have to be in a C-odd state.  However, C invariance is violated by the
mass-splitting between $L$ and $R$-chiral squarks (superpartners of
the left and right chiral quarks).  This mass splitting is a model
dependent aspect of SUSY breaking.  Fortunately, as we will see, our
analysis is quite insensitive to the extent of C violation.

We now present expressions for the transition rates to be used in
\eqr{eq:boltz3}: $\avg{W_{\pho \pi \rightarrow \r0} V} n^{eq}_{\pi}$,
$\avg{W_{\pho \pi \rightarrow \pi \r0} V} n^{eq}_{\pi}$, $\avg{W_{\pho
\pi \pi \rightarrow \r0} V^2} n^{eq}_{\pi} n^{eq}_{\pi}$, $\avg{W_{\r0
\r0 \rightarrow X} V} n_\r0^{eq}$, and $\avg{W_{\pho \pho \rightarrow X}V}
n_\pho^{eq}$.  We shall see that the resulting
expressions do not differ significantly from those of Ref.\
\re{farkolb} even though the issue of charge conjugation symmetry is
ignored in that reference.

\vspace{24pt}
\centerline{A. The \intcvt reaction $\pho \pi \rightarrow \pi \r0$}  

If charge conjugation invariance were exact, the neutral pion channel
would be absent as it necessarily violates C.  However, even if C is
maximally violated, the condition $\sigma(\pho \pi^{0} \rightarrow
\pi^{0} \r0) \ll \sigma(\pho \pi^{\pm} \rightarrow \pi^{\pm} \r0)$
still applies because the $\pho \r0 q \overline{q}$ coupling is
proportional to the quark charge, causing a first order cancelation
to occur in the case of a neutral $\pi^0$.  Thus we can ignore the
neutral channel without serious impact on the quantitative results.
In order to avoid any thermal averaging complications that may arise
from threshold effects, we estimate the cross section for $\r0 \pi^{\pm}
\rightarrow \pi^{\pm} \pho$ instead of the cross section for the
inverse reaction.  The cross section formula is the same as
that given in \re{farkolb}: 
\be
\langle W_{\r0 \pi^{\pm} \rightarrow \pi^{\pm} \pho} V\rangle=
\avg{ v\sigma_{\r0\pi} } \simeq  1.5\times10^{-10}\, r  \,
\left[\mu_8^2 \mu_{S}^{-4}C\right] \ \mb  .
\label{eq:EQA}
\ee The factor $C$ contains the uncertainty due to possible resonance
effects and hadronic physics.  Ref.\ \re{farkolb} considered the range
$1 \le C \le 10^3$.  An analysis of the effect of the expected $\rpi$
resonance shows that $C$ can exceed this by an order of magnitude (see
Appendix), but we shall not dwell on this since our conclusions are
mostly insensitive to the exact value of any large enhancement.
However, for $C/\mu_S^4 \la 1$ the results are sensitive to the value
of $C/\mu_S^4$.  For reasons to be discussed in Section V, we also
consider values of $C$ as small as 1/20.

Using \eqr{eq:timerev} and $n_j^{eq} \approx g_j (m_j T /(2
\pi))^{3/2} e^{-m_j/T}$, we then find
\begin{eqnarray}
\langle W_{\pho \pi^{\pm} \rightarrow \pi^{\pm} \r0} V\rangle
n^{eq}_{\pi^\pm} & = & \frac{n_\r0^{eq}}{n_\pho^{eq}}
                n_{\pi^\pm}^{eq}  \avg{v\sigma_{\r0\pi}} \nonumber \\
& = & 9.17\times10^{-13}\ r^{5/2}\ x^{-3/2}\ \exp(-0.175\mu_8^{-1}x)
\GeV
\nonumber
\\
& & \times  \exp[-(r-1)x]\ [\mu_8^{7/2}\mu_S^{-4}C]  \ .
\label{eq:interconvert}
\end{eqnarray}
Note in \eqr{eq:interconvert} that parameters $C$ and $\mu_S$
occur only together in the combination $C/\mu_S^4$, which we take to
lie in the range
\be
6.17 \times 10^{-4} \la \frac{C}{\mu_S^4} \la  10^4
\label{eq:endpt1}
\ee
in accordance with the limit on $\mu_S$ given by Table I and
\eqr{mparams}.

\vspace{24pt}
\centerline{B. The inverse decay reactions} 

We now estimate the decay rate of $\r0$ and use \eqr{eq:timerev}
to obtain the inverse decay rate.  If charge conjugation invariance is
exact, two body decays of an $\r0$ to a $ \pho $ and a pseudoscalar
meson (C=+1) are forbidden [\ref{fPRL}].
In order to avoid reliance on a model of SUSY-breaking and its
predictions for the extent of C-violation, we parameterize the
branching fraction of an $R^0$ to 2- and 3-body final states by $b_2$
and $b_3$, respectively.  As in the \intcvt reaction, the neutral pion
channel ($\r0
\rightarrow \pho \pi^0 \pi^0$)  can
be safely ignored even if $b_3 \gg b_2$.  When $b_2$ is not negligible,
the two-body final states could be $\pho \pi^0$ and $\pho \eta$.
However, the matrix element-squared for $\r0 \leftrightarrow \eta \pho
$ is about one-quarter of that for $\r0 \leftrightarrow \pi^0 \pho$
[\ref{fPRL}], and the $\eta$ final state is additionally suppressed by
phase space.  Hence we make an unimportant error by retaining only the
two-body final state $\pho \pi^0$.  Thus, the two reactions of
interest are $\r0 \rightarrow \pho \pi^+ \pi^-$ and $\r0 \rightarrow
\pho \pi^0 $, with branching fractions $b_3$ and $b_2$,
respectively.  In this subsection, we show that our results depend
only minimally upon the individual magnitudes of $b_2$ and $b_3$
because the Boltzmann equation depends only on the total decay width
of the $\r0$ and $b_2+ b_3 \approx 1$ (due to the relative phase space
suppression of 4-body decays).

The rates $\Gamma(\r0 \rightarrow \pho \pi^0)$ and $\Gamma(\r0
\rightarrow \pho \pi^+ \pi^-)$ are obtained from the $\r0$ decay rate in Ref.\
\re{farkolb} by inserting $b_2$ and $b_3$ to get
\be
\label{eq:FR}
W_{\r0\rightarrow\pho\pi^+\pi^-}=
\Gamma_{\r0\rightarrow\pho\pi^+\pi^-} = 2.0\times10^{-14} \ {\cal
F}(r) \theta(r-0.350 \mu_8^{-1} -1) \ \GeV\ [\mu_8^5\mu_S^{-4}B b_3] \
\ee
and 
\be
\label{eq:FR2}
W_{\r0\rightarrow\pho\pi^0}= \Gamma_{\r0\rightarrow\pho\pi^0} =
            2.0\times10^{-14} \ {\cal F}(r)\theta(r-0.175
\mu_8^{-1} -1)\ \GeV\ 
[\mu_8^5\mu_S^{-4}B b_2] \ 
\ee
where ${\cal F}(r)=r^5(1 - r^{-1})^6$, $\theta$ is a step function
employed to model the threshold of the decay channel, and the factor
$B$ reflects the overall uncertainty which we set to be in the range
$1/300 \la B \la 3$.  Having obtained the decay rate formula, we now
use \eqr{eq:timerev} to convert it to the inverse decay rate\footnote{
A more accurate relationship between the thermal averaged decay rate
and the non-thermal averaged decay rate is
$\langle\Gamma_{\r0\rightarrow\pho\pi^+\pi^-}\rangle=
\Gamma_{\r0\rightarrow\pho\pi^+\pi^-} K_{1}(r x)/K_{2}(r x)$ where
$K_{\nu}$ is the modified Bessel function irregular at the origin.
Since the freeze out occurs typically between $x=20$ and $x=30$, the
thermal averaged reaction rate will maximally deviate from the
non-thermal averaged one when $r=1.1$ and $x=20$.  In that case
$K_{1}(r x)/K_{2}(r x)= 0.94$ which is still an insignificant
correction.  Thus we neglect this complication in our calculations.}
\begin{eqnarray}
\lefteqn{\langle W_{\pho \pi^+ \pi^- \rightarrow \r0} V^2 \rangle n_{\pi^+}^{eq} n_{\pi^-}^{eq} =
\langle\Gamma_{\r0\rightarrow\pho\pi^+\pi^-}\rangle \frac{n_\r0^{eq}}{n_\pho^{eq}}} \\
& = & 2.0\times10^{-14} r^{3/2} {\cal F}(r) e^{-(r-1)x} \theta(r-0.350
\mu_8^{-1} -1)  \GeV
[\mu_8^5\mu_S^{-4}B b_3],
\label{eq:inversedec}
\end{eqnarray}
and similarly 
\begin{eqnarray}
\lefteqn{\langle W_{\pho \pi^0 \rightarrow \r0} V \rangle n_{\pi^0}^{eq} =
\langle\Gamma_{\r0\rightarrow\pho\pi^0}\rangle
\frac{n_\r0^{eq}}{n_\pho^{eq}} } \\
& =&  
2.0\times10^{-14} r^{3/2} {\cal F}(r) e^{-(r-1)x} \theta(r-0.175
\mu_8^{-1} -1) \GeV
[\mu_8^5\mu_S^{-4}B b_2].
\label{eq:inversedec2}
\end{eqnarray}
Combining Eqs.\ (\ref{eq:inversedec}) and (\ref{eq:inversedec2}) with
\eqr{eq:R_tot}, and using $b_2+b_3 \approx 1$, we find 
\be
R_{tot} = \Gamma_{tot} g(b_2,r, \mu_8)  + 2 \avg{W_{\pho \pi^{\pm}
\rightarrow \pi^{\pm} \r0} V} n^{eq}_{\pi^\pm}
\label{eq:rtotsimp} 
\ee
where $\Gamma_{tot} \equiv 2.0\times10^{-14} r^{3/2} {\cal F}(r)
e^{-(r-1)x} \GeV [\mu_8^5\mu_S^{-4}B ]$ and
\[
g(b_2, r, \mu_8)\equiv \left\{ \begin{array}{ll}
			  1 & \mbox{if $r > 0.35 \mu_8^{-1} +1$} \\
			b_2 & \mbox{if $ 0.35 \mu_8^{-1} +1 \geq r >
0.175 \mu_8^{-1}+1$} \\
0 & \mbox{otherwise}
\end{array}
\right. .
\]
The function $g$ allows both the two and the three body decays when
the $\r0$ is sufficiently heavy ($r > 0.35 \mu_8^{-1} +1$) but forbids
the three body channel when the $\r0$ mass drops below the two pion
channel threshold.  Thus, as long as the parameterization in Eqs.\
(\ref{eq:inversedec}) and (\ref{eq:inversedec2}) is valid and the
$\r0$ is massive enough ($r > 0.35 \mu_8^{-1} +1$) to allow
kinematically three body decays, our results are independent of $b_2$
and $b_3$, and hence the question of C invariance.  Therefore,
considerations of C invariance is generally unimportant for large
values of $r$.

In these formulae, the squark mass parameter $\mu_S$ occurs only in
combination with the uncertainty parameter $B$ in the form
$B/\mu_S^4$.  Using Table I and Eq.\ (\ref{mparams}), we limit the
decay rate for a given $\mu_8$ and $r$ to those values corresponding
to the range\footnote{Because of our estimated upper and lower limit
on each of the parameters $C, B$, and $\mu_S$ separately
(\eqr{eq:endpt1},
\eqr{eq:endpt2}, and Table I), for a given value of
$B/\mu_S^4$, the allowed range of values for $C/\mu_S^4$ given by
\eqr{eq:endpt1} must be supplemented with the condition
\be
\left[\frac{C_{min}}{B_{max}}\right]\frac{B}{\mu_S^4} \la \frac{C}{\mu_S^4} \la
\left[\frac{C_{max}}{B_{min}}\right]\frac{B}{\mu_S^4}
\label{eq:cuts}
\ee
where from \eqr{eq:endpt1} $C_{max} = 1000$,
$C_{min}=1/20$, and $B_{max}=3$, and $B_{min}=1/300$.}
\be 4.12 \times 10^{-5} \la \frac{B}{\mu_S^4} \la 48.
\label{eq:endpt2}
\ee

\vspace{24pt}
\centerline{C. Self-Annihilations and co-annihilations}

For the thermal averaged $\r0$ self-annihilation cross section, we use
$\avg{v \sigma_{\r0 \r0}} = 31 A \mbox{ mb}$.  This is extracted from
the $p \bar{p}$ annihilation cross section in the comparable kinematic
region\re{agnello} with a
factor $A$ inserted to cover a possible difference between $\r0 \r0$
and $p \bar{p}$ annihilation, and to account for the uncertainty due
to possible resonance enhancements and other hadronic effects.  We
take $A$ to lie in the range $10^{-2} \la A \la 10^2$.\footnote{Note
the absence of the $v^2$ factor which appears in the familiar case of
two identical Majorana spinors annihilating to a fermion-antifermion
pair (e.g., $\pho \pho$ or $\glu \glu \rightarrow q \bar{q}$).  Like the
$\pho \pho$ and $\glu \glu$ states, the $\r0 \r0$ system must be
antisymmetric by Fermi statistics, i.e., $^1S_0,~ ^3P_1,...$.  However
typical final states of $\r0 \r0$ annihilation (e.g., 3 pions) can
have $0^{-+}$ quantum numbers, allowing $s$-wave annihilation.  This
is to be contrasted with the usual case that the final state is a
fermion-antifermion pair.  Since the sfermion-fermion-gaugino
interaction conserves chirality, the $0^{-+}$ state in that case is
helicity-suppressed and thus $p$-wave annihilation is necessary.  This
treatment departs from Ref.\ \re{farkolb}, but does not lead to
significantly different conclusions than the $\avg{v \sigma_{\r0 \r0}} =
100 A v^2 \mbox{ mb}$ used there.}  
Hence, the $\r0$ self-annihilation rate is given by 
\be
\avg{W_{\r0 \r0 \rightarrow X}V} n_\r0^{eq}=10 A r^{3/2} x^{-3/2}
\exp(-rx) [\mu_8^3] \GeV.
\label{eq:r0ann}
\ee
Note that although the $\r0$ self-annihilation rate is generally much
larger than the other reaction rates before the $\pho$ freeze out
time, it is usually not strong enough to maintain $\r0$ in equilibrium
abundance through the $\pho$ freeze out time.  This fact, not taken
into account in Ref.\ \re{farkolb}, leads to differences
in the results between that paper and the present analysis.

The well known thermal average of the $\pho$ self-annihilation cross
section [\ref{G},\ref{LSP},\ref{INDIRECTA},\ref{KKC}] can be
approximated as \re{farkolb} $\avg{ v \sigma_{\pho \pho}}= 2.0 \times
10^{-11} x^{-1} [
\mu_8^2 \mu_S^4]
\mbox{ mb}$ for our purposes, giving the transition rate
\be
\avg{W_{\pho \pho \rightarrow X} V} n_\pho^{eq} =3.3 \times 10^{-12}
x^{-5/2} \exp(-x) [\mu_8^5 \mu_S^{-4}] \GeV.
\ee
Because the $\pho$ self-annihilation becomes ineffective earlier than
the $\r0$ self-annihilation, it contributes very little to our results.

In summary, the reactions that will be important to our system of
equations are $\r0 \r0 \leftrightarrow X$, $\r0 \pi \leftrightarrow
\pho \pi$, $\r0 \leftrightarrow \pho \pi \pi$, and $\r0
\leftrightarrow \pho \pi$.

\vspace{36pt}
\centerline{\bf IV. GENERAL RESULTS}
\vspace{24pt}

In this section, we impose the cosmological constraint $\Omega_\pho
h^2 \leq 1$ on the integration results of the Boltzmann equation to
identify the allowed region of the parameter space and use the
condition $\Omega_\pho h^2 \geq 0.01$ to identify those parameters for
which the photinos are significant dark matter candidates.
 The
parameter space is spanned by $r\equiv M/m, B/\mu_S^4, C/\mu_S^4,
\mu_8,$ and $A$ (see Table I and Eqs.\ (\ref{mparams}),
(\ref{eq:EQA}), (\ref{eq:FR}), and (\ref{eq:r0ann})).  For reasons of
physical interest, we will present our results in terms of the $\pho$
mass $m$ and the $\r0$ mass $M$ instead of using $r$ and $\mu_8$.  We
constrain the parameter space for the two extreme cases $b_2=1$ and
$b_2=0$ (maximal C violation and C conservation, respectively), but in
general, the results are insensitive to the value of $b_2$.  As will
be discussed below, among the parameters of the model, the relic
abundance is most sensitive to the variations of $r$.  Using 
a maximum $A$ of about 100, our analysis gives us an upper bound of $r
\le 1.8$.

\begin{figure}
\hspace*{25pt} \epsfxsize=400pt \epsfbox{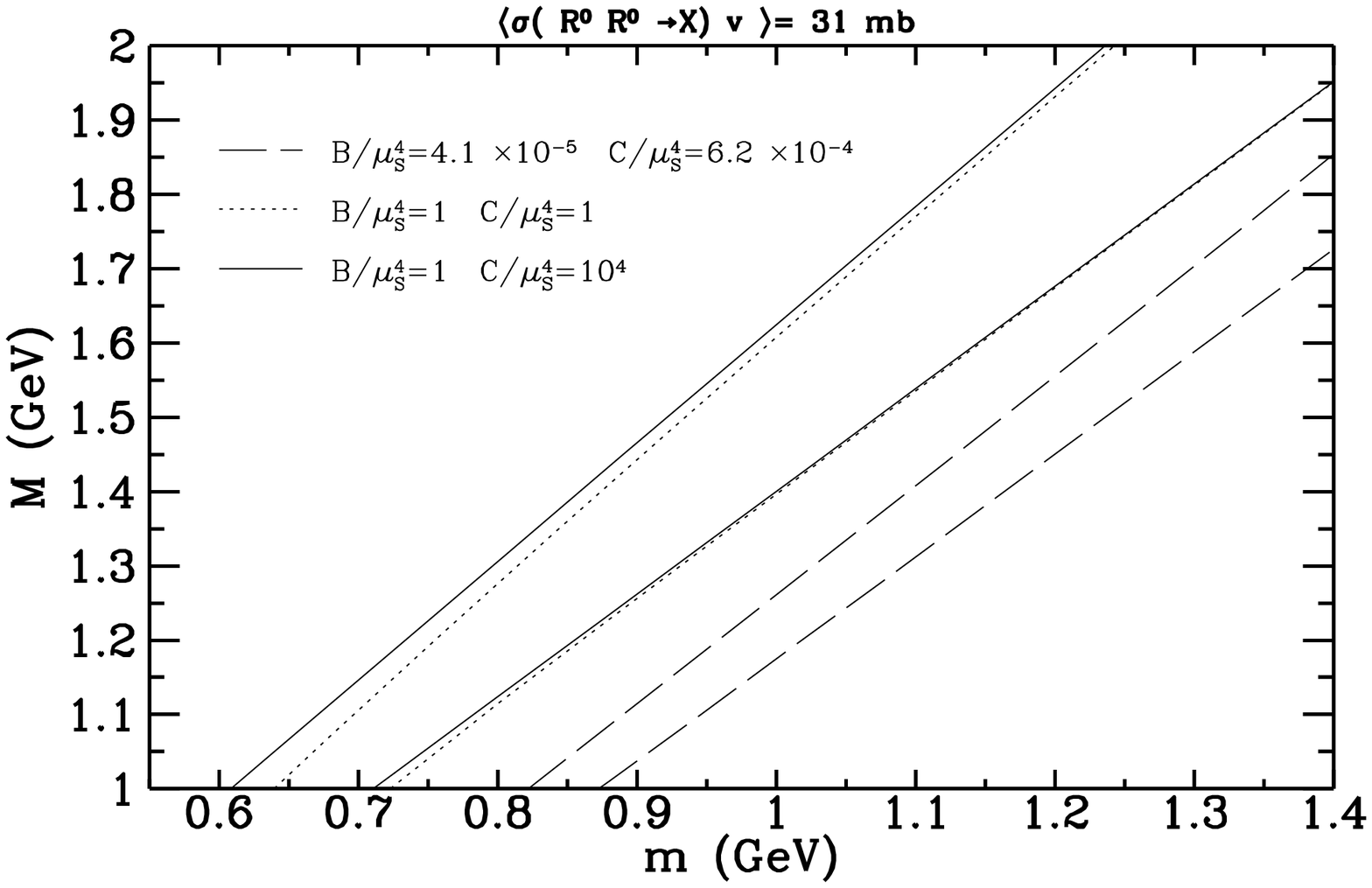}\\
\hspace*{1em} {\footnotesize{{\bf Fig.\ 1:} For any given contour type,
the left contour gives those values of $M$ ($\r0$ mass) and the $m$
($\pho$ mass) for which $\Omega_\pho h^2$=1 while the right one gives
those for which $\Omega_\pho h^2=0.01$.  The region above the left
contour is ruled out by the present analysis.} }
\end{figure}

\begin{figure}[p]
%
\hspace*{25pt} \epsfxsize=400pt \epsfbox{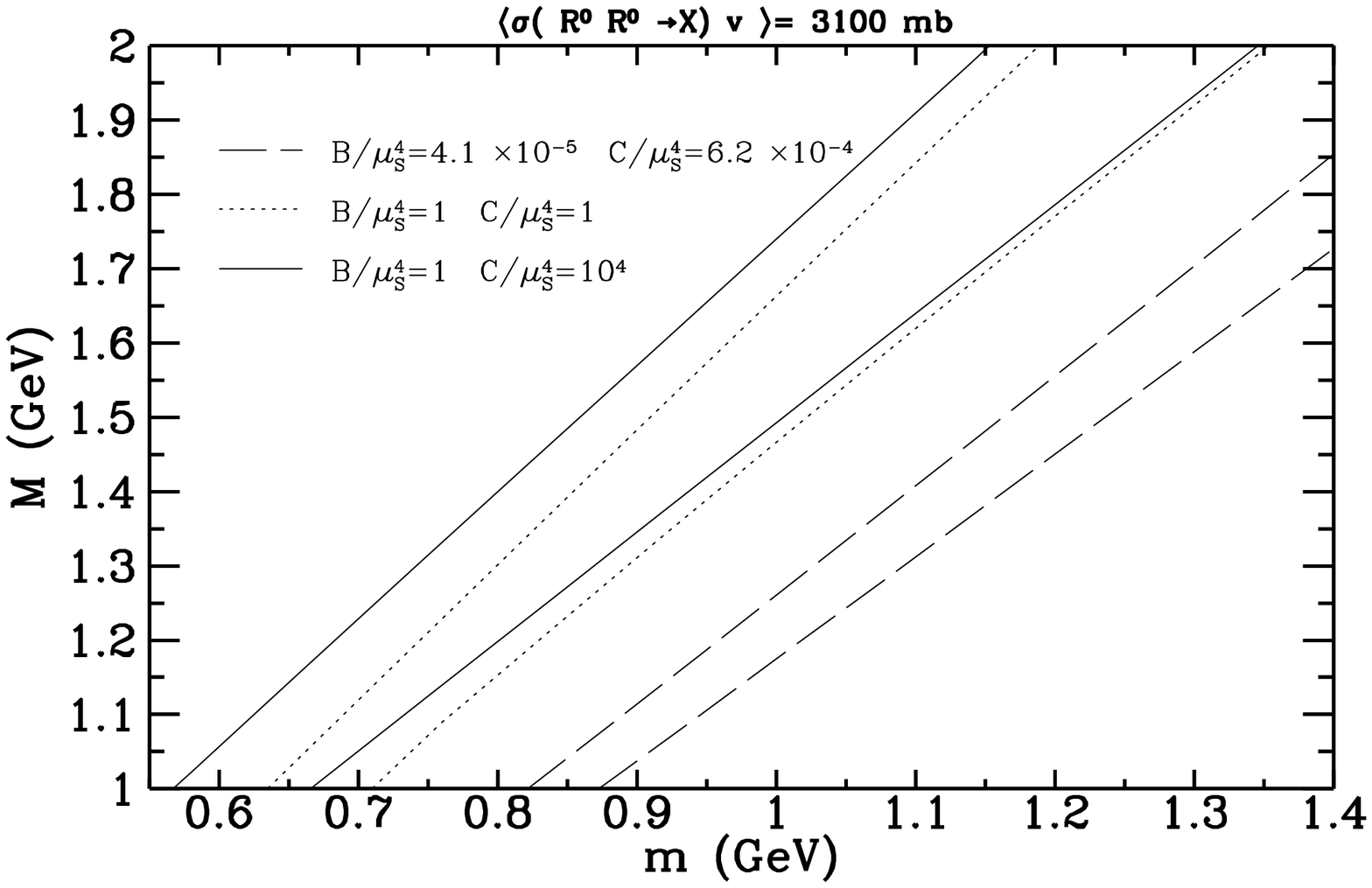}\\
\hspace*{25pt} \epsfxsize=400pt \epsfbox{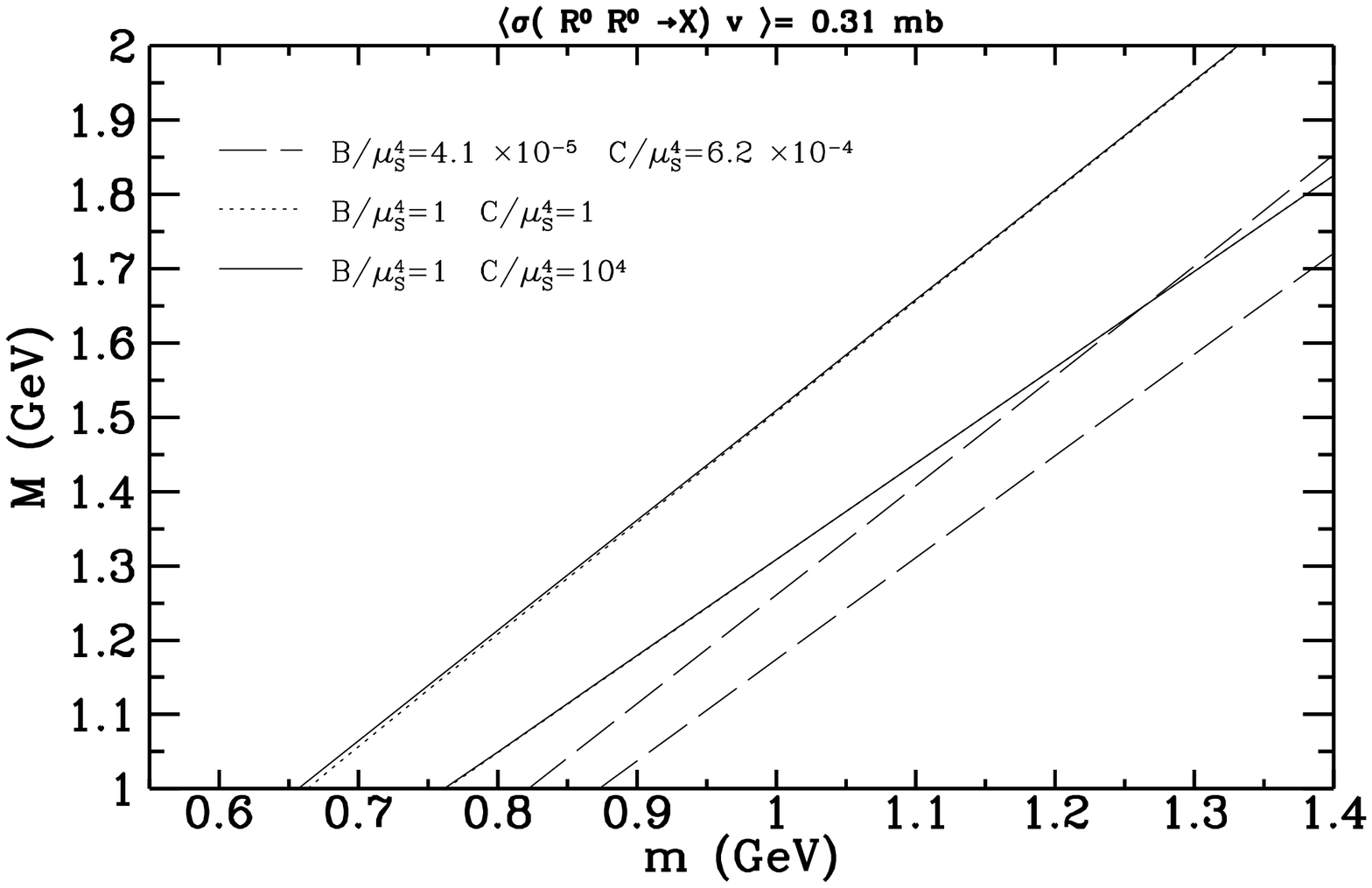}\\
\hspace*{1em} {\footnotesize{{\bf Fig.\ 2:}  Same as Fig.\ 1 except
for the $\r0$ self-annihilation cross sections.  For the top figure,
$\avg{v \sigma(\r0 \r0\rightarrow X) } = 3100$ mb while for
the bottom figure, $\avg{v \sigma(\r0 \r0\rightarrow X) } = 0.31$ mb.}}
\end{figure}

In Fig.\ 1, we show the Boltzmann equation integration results with
exact C invariance ($b_2=0$).  For any given contour type, the left
contour represents the $\Omega_\pho h^2=1$ contour while the right
contour represents the $\Omega_\pho h^2=0.01$ contour.  The present
analysis thus excludes the region above the left contour and
constrains the masses to lie between a given contour type in order for
the $\pho$ to be a significant source of dark matter (defined by
$\Omega_\pho h^2 \geq 0.01$).  In this figure, the parameter $A$
multiplying the $\r0$ self-annihilation cross section has been set to
1.

Note that the values of $r \equiv M/m$ are insensitive to $C/\mu_S^4
\ga 1$.  This can be heuristically understood by the fact that as
$C/\mu_S^4$ increases, the freeze out time (the time at which the
\intcvt reaction rate becomes negligible compared to $H$ if the \intcvt
rate dominates over the inverse decay rate) approaches the time at
which the $\r0$ self-annihilation rate becomes negligible compared to
$H$.  Thus, as $C/\mu_S^4$ increases, the photino abundance should
approach the value for the limiting case when the $\r0$
self-annihilation rate becomes negligible {\em before} the freeze out
time.  When the $\r0$ self-annihilation rate becomes negligible, the
number of SUSY particles are approximately conserved.\footnote{The
$\pho$ self-annihilation rate is already negligible by the the time
the $\r0$ self-annihilation becomes negligible.} Thus, the $\pho$
abundance is largely determined by the time at which $\r0$
self-annihilation becomes negligible in this limiting case.  This time
is determined by $r$ and is independent of $C/\mu_S^4$.\footnote{This
heuristic argument assumes that the $\r0$ and $\pho$ abundances
approximately follow a function independent of $C/\mu_S^4$ until near
the time that the SUSY particles become approximately conserved.  This
is of course true for the equilibrium abundance functions.}

Because the extent of C violation affects the photino abundance only
in the region $0.14 \GeV \leq M-m \leq 0.28 \GeV$, it is clear from
Fig.\ 1 that only the long dashed contours (corresponding to small
\intcvt and inverse decay rates) may depend on the extent of C
violation.  However, for this region, $B/\mu_S^4$ is too small for the
inverse decay reaction to play any significant role, and hence, our
results are insensitive to the extent of C violation.  Explicit
numerical calculations confirm this.

In Fig.\ 2, we show the effect of changing the magnitude of the $\r0$
self-annihilation cross section (by changing $A$ in Eq.\
(\ref{eq:r0ann})).  When we increase the magnitude from that of Fig.\
1 by a factor of 100 (due to a possible resonance enhancement), the
contours for $C/\mu_S^4, B/\mu_S^4 \geq 1$ shift leftwards, and when
we decrease the magnitude by a factor of 100, the same contours shift
rightwards.  In both cases, the contours corresponding to a small
$C/\mu_S^4$ and $B/\mu_S^4$ (corresponding to small \intcvt and
inverse decay rates) remain essentially unchanged.  This is expected
since for the shifted contours, the inverse decay and the
\intcvt reaction rates are large enough such that the $\pho$ abundance
is sensitive to the time at which the $\r0$ self-annihilation becomes
negligible (by the mechanism discussed before) while for the
unchanging contours, the $\pho$ abundance is determined nearly
independently of the time that $\r0$ self-annihilation rate becomes
negligible.  When the inverse decay and the \intcvt reaction rates are
very small as is the case for the unchanged contours, the $\pho$
freeze out time
that will lead to $\Omega_\pho h^2 \ga 0.01$ is much earlier than the
time when the $\r0$ self-annihilation reaction rates become
negligible.  Hence, near the $\pho$ freeze out time, the $\r0$
self-annihilation reaction rate will dominate the $dY_\r0/dx$, the
$\r0$ abundance will be nearly in equilibrium, and the $dY_\pho/dx$
will decouple from $dY_\r0/dx$, leading to a $\pho$ freeze out value
that is nearly independent of the $\r0$ self-annihilation reaction.
Note also that when the time at which the $\r0$ self-annihilation
becomes negligible is pushed away from the $\pho$ freeze out time by
increasing the $\r0$ self-annihilation cross section, the solid and
the dotted contours become more sensitive to the value of $C/\mu_S^4$
as we expect from our heuristic discussion above.

According to Fig.\ 2, the maximum value of $r$ allowed by the
condition that $\Omega_\pho h^2 \leq 1$ is about 1.8.  


\vspace{36pt}
\centerline{\bf V. CROSSING RELATION}
\vspace{24pt}

The amplitudes determining the quantities $\avg{v \sigma} \equiv
\avg{v \sigma_{\r0 \pi^{\pm} \rightarrow \pho \pi^{\pm}}}$ and
$\Gamma_{tot}$ are related through crossing symmetry if we associate
$\Gamma_{tot}$ with the C conserving $\r0 \rightarrow \pho \pi^+
\pi^-$ transition rate (i.e., if we set $b_2 =0$).  To obtain a useful
constraint from the crossing relation, and to implement the
constraints following from the symmetries of the underlying theory, we
derive in this section an approximate effective interaction
Lagrangian.  If the $\rpi$ resonance is sufficiently far above
threshold such that the $\r0 \pi^{\pm} \rightarrow \pho \pi^{\pm}$
amplitude can be taken to be momentum independent for the purposes of
the freeze-out calculation, a single parameter governs both $\avg{v
\sigma_{\r0 \pi^{\pm} \rightarrow \pho \pi^{\pm}}}$ and
$\Gamma_{tot}$.  This allows us to determine what ranges of $R^0$
lifetime are most favorable for cosmology, in the event the $\rpi$ is
too far above threshold to have a significant impact.

We first note that neglecting light quark masses as well as left-right 
squark-mass splitting in comparison to the squark masses, the four-Fermi
effective operator governing $R^0 \pi \leftrightarrow \pho \pi$ can be
written in the current-current form
\begin{equation}
 {\cal H}_{\rm int} = i\kappa_V\ \lambda_\glu^a \, \gamma^\mu \lambda_\pho\
\bar{q}^i \gamma_\mu \, T^a_{ij} q^j + 
\kappa_A\ \lambda_\glu^a \,  \gamma^\mu \gamma_5 \lambda_\pho\ 
\bar{q}^i \gamma_\mu \gamma_5 \, T^a_{ij} q^j
\label{L4q}
\end{equation}
where $\lambda_\glu$ and $\lambda_\pho$ are 4-component Majorana
spinor fields for the gluino and photino, $\{a,\ i,\ j\}$ are color
indices, and the $ T^a$ are $3 \times 3$ SU(3) matrices.  This form
follows because the underlying theory conserves the chirality of light
quarks and their SUSY partners,\footnote{Chirality
conservation of the light quarks and their squarks is an excellent
approximation in all SUSY models proposed to date, for which
left-right squark mixing is proportional to the mass of the
corresponding quark.} allowing only current-current couplings for the
quarks to appear.  Approximate degeneracy of the left-right squark
masses then ensures parity conservation which, with Lorentz
invariance, results in the form of Eq.\ (\ref{L4q}).  A direct
calculation starting with the fundamental supersymmetric Lagrangian of
course gives the form (\ref{L4q}) and gives $\kappa_V = 0$ and $\kappa_A =
g_Se_q e/\mst^2$.\footnote{The strong coupling constant is
denoted by $g_S$ and $e_q$ gives the electric charge of the quark in
units of positron charge $e$.}  The vanishing of $\kappa_V$ is due to
C-conservation, since the term it multiplies is C-odd for Majorana
fields $\lambda_\glu$ and $\lambda_\pho$.

We are concerned with estimating matrix elements such as $\avg{R^0 \pi
| {\cal H}_{\rm int}|\pho \pi}$.  The most general form of the matrix
element includes current-current terms, plus other terms which result
from the fact that the $R^0$ is not pointlike and chirality flip can
be induced by long-distance effects.  However, since the $R^0$ is
expected to be more compact than ordinary hadrons (as is observed for
the $0^{++}$ glueball\footnote{D. Weingarten, private
communication.}), we neglect all but the current-current terms.
Therefore we have ${\cal L}_{\rm eff} = i \kappa \overline{R^0}
\gamma^\mu \lambda_\pho J_\mu$, where $J_\mu$ is a
C-odd,\footnote{Because the $\r0$ and $\pho$ have opposite C quantum
numbers [\ref{fPRL}].} four-vector pion current determined by chiral 
perturbation theory, and $\kappa$ is of order $\kappa_A$.  The
single-pion contribution to $J_\mu$ vanishes, and the two pion
contribution is simply $ J_\mu = i(\pi^\dagger \partial_\mu \pi -
(\partial_\mu \pi)^\dagger \pi)$.  In general, $\kappa$ is a function
of kinematic invariants, but far from resonances a constant should be
a reasonable approximation.

\begin{figure}[p]
%
\hspace*{25pt} \epsfxsize=400pt \epsfbox{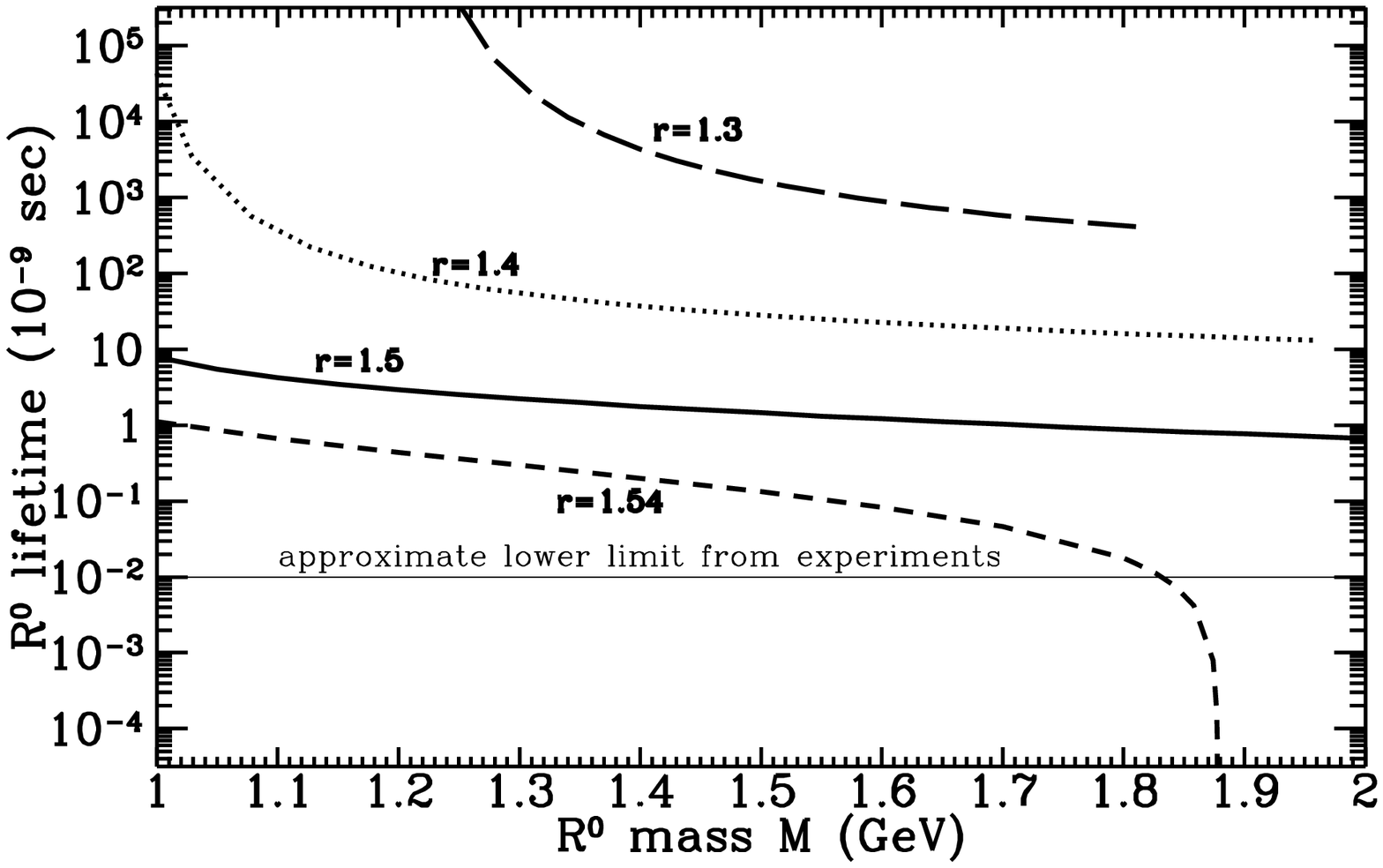}\\
\hspace*{1em} {\footnotesize{{\bf Fig.\ 3:} The $\r0$ lifetime that
implies a cosmological photino abundance of $\Omega_\pho h^2= 0.25$ is
plotted as a function of the $\r0$ mass and its ratio $r$ to the
photino mass.  A model Lagrangian has been used to determine the
crossing relation between the \intcvt amplitude and the $\r0$ decay
amplitude. }}

%
\hspace*{25pt} \epsfxsize=400pt \epsfbox{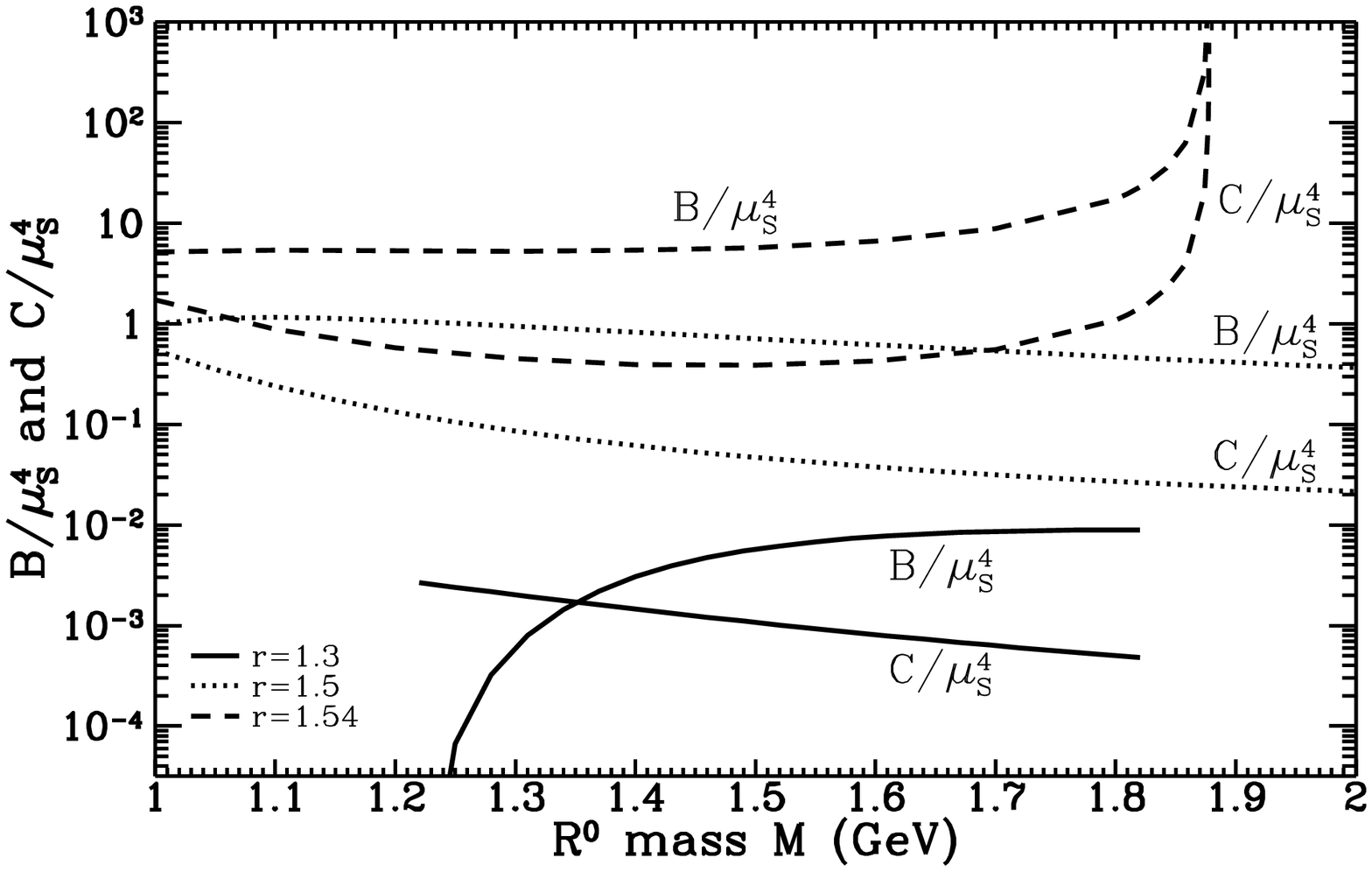}\\
\hspace*{1em} {\footnotesize{{\bf Fig.\ 4:} The $B/\mu_S^4$ and
$C/\mu_S^4$ values corresponding to the contours shown in Fig.\ 3 are
plotted.  The typical suppression of $C/\mu_S^4$ with respect to
$B/\mu_S^4$ reflects the fact that both the \intcvt and the inverse
decay reactions have comparable rates in our simple model which does
not take into account possible resonance enhancements.}}
\end{figure}

Using ${\cal L}_{\rm eff} $ we can compute both $\avg{v \sigma} \equiv
\avg{v \sigma_{\r0 \pi^{\pm} \rightarrow \pho \pi^{\pm}}}$ and
$\Gamma_{tot}$ in terms of the single parameter $\kappa$.  Thus, for a
given $r$ and $M$, $\Omega_\pho h^2$ is a function of the single
parameter $\kappa$.  Likewise, values for $\{\Omega_\pho h^2$, $r$,
$M$\} pick out a unique value of $\kappa$, which in turn determines
$\Gamma_{tot}$.  In Fig.\ 3, we assume $\Omega_\pho h^2 = 0.25$
(cosmologically ``favored'' value) and give the $\r0$ lifetime for a
range of $r$ and $R^0$ mass.  We stress that these results are only
indicative of the actual lifetime-mass-relic density relation, since
the most general effective Lagrangian depends on additional parameters
which we neglect here.  Furthermore if the $\rpi$ resonance is
sufficiently close to threshold to produce an enhancement effect,
there is no simple relation among $\Omega_\pho h^2$, $r$, and the
$\r0$ lifetime, and Fig.\ 3 is not relevant.  It is encouraging that
the $R^0$ lifetimes required to give the ``correct'' relic density is
compatible with predictions \re{PHENO} and also compatible with
experimental limits \re{EXPTS}.

It is also of interest to extract the values of $B$ and $C$ implied by
$\Omega_\pho h^2 = 0.25$; this is shown in Fig.\ 4.  Overall, these
results suggest that the inverse decay reaction may not be entirely
negligible in determining the photino abundance if there is no
resonance enhancement in the scattering reaction.


\vspace{36pt}
\centerline{\bf VII. SUMMARY AND CONCLUSIONS}
\vspace{24pt}
In this article, we have investigated the cosmological constraints on
the physics of light photinos and gluinos.  A full treatment of the
Boltzmann equations governing the photino freeze out has been carried
out, considering the total $\r0$ width and $\r0 \pi \rightarrow \pho
\pi$ scattering cross section as independent quantities.  We find that 
to avoid photino abundances inconsistent with cosmology, the ratio $r$
of $\r0$ mass to $\pho$ mass must be less than about $1.8$.
We checked that if the $\r0$ is the LSP, its annihilation is too
efficient for it to account for the observed dark matter
density.

We also developed an approximate effective Lagrangian description of
the $R^0 \pi \leftrightarrow \pho \pi$ amplitude, neglecting possible
C-violating and chirality-violating effects.  If the $\rpi$
resonance is far above threshold, ${\cal L}_{\rm eff}$ is specified by a
single parameter governing both the total $\r0$ width and $\r0 \pi
\rightarrow \pho \pi$ scattering cross section.  Assuming that 
the universe is at its critical density with photinos constituting
most of the dark matter fixes this parameter for given $\r0$ and
$\pho$ masses.  We therefore obtain the cosmologically favored
lifetime range of the $R^0$ as a function of its mass (shown in Fig.
3) in the absence of a low lying $R_{\pi}$ resonance.  
The lifetime will be increased compared to the values given in Fig. 3
when the $\rpi$ resonance enhances the cosmological importance of the
scattering cross section in comparison to the inverse decay.  
Although the limitations in this estimate must not be forgotten, it is
encouraging that the range thus determined, $ \tau > 10^{-10}$s, is
compatible with experimental limits \re{EXPTS}.  Much of this range of
lifetimes should be accessible to direct observation in upcoming
experiments \re{fPRL}.

In closing, we note that detectability of relic dark matter is
different for light $\pho$'s than in the conventional heavy WIMP
scenario for two reasons.  Firstly, the usual relation between the
relic density and the WIMP-matter scattering cross section only
applies when the relic density is determined by the WIMP
self-annihilation cross section, whereas in the light photino scenario
it is determined by the \intcvt cross section,
$\r0$ self-annihilation cross section, and the $\pi^\pm$ density at
freeze out.  Secondly, WIMP detectors have generally been optimized to
maximize the recoil energy for a WIMP mass of order $10$ to $100$ GeV.
Goodman and Witten in Ref.\
\re{goodwitt} discuss $\pho$ detection through $\pho$-nucleon elastic
scattering. Using Eq.\ (3) of Ref.\ \re{goodwitt} and the parameters
discussed here, one finds that event rates range somewhere between
$10^{-3}$ and $10$ events/(kg day).  Unfortunately even if the event
rate were larger, observation of relic light photinos would be
difficult with existing detectors because the sensitivity of a generic
detector is poor for the less than $1$ GeV mass relevant in this case.

\vspace{36pt}
\centerline{\bf ACKNOWLEDGMENTS}
\vspace{24pt}
DJHC and EWK were supported by the DOE and NASA under Grant NAG5-2788.
GRF was supported by the NSF (NSF-PHY-94-2302).

\vspace{36pt}
\centerline{\bf APPENDIX}
\vspace{24pt}
In this Appendix, we give the formalism for treating the resonance
enhancement of the $\r0 \pi \rightarrow \pho \pi$ cross section, using
a Breit-Wigner form for the resonance.  This permits us to assess the
plausibility of the original range used in Ref.\ \re{farkolb}, $C <
1000$.  We find that the effective value of $C$ could be significantly
larger than the originally estimated upper bound, but this is only
relevant if $\avg{v \sigma_{\r0 \r0}}$ is large enough that $\r0$'s
remain in thermal equilibrium until after photino freeze out.  As
discussed in Section IV, this is not the case for large $C$, given our
estimated $\avg{v \sigma_{\r0 \r0}}$.  However if there were a
$0^{-+}$ glueball near $\r0 \r0$ threshold, the $\r0$'s could stay in
equilibrium to a lower temperature, and make it necessary to include
resonance effects for both self-annihilation and \intcvt processes.
We treat below the modeling of a resonance in the $\r0 \pi \rightarrow
\pho \pi$ reaction; the treatment of a resonance in the $\r0 \r0$
self-annihilation cross section is a close parallel.

The resonance relevant to the $\r0 \pi \rightarrow \pho \pi$ reaction
is called $\rpi$ which is composed at the valence level of $\glu$,
$q_1$, and $\overline{q}_2$ (where $q_i$'s are $u$ and $d$ quarks).  To
study the maximum enhancement, we consider the $\rpi$ mass to be close
to the $\r0$ mass.  We also consider here only the charged $\rpi$'s
since we are concerned with charged pion scattering (see Section
III).  Furthermore, because the $s$-wave contribution 
dominates, we restrict ourselves to the $J=1/2$ state.

We write the resonant contribution to the $\r0 \pi
\rightarrow \pho \pi$ cross section as
\be
\sigma_{\mbox{res}}= \left[\frac{4 \pi}{p_{cm}^2}\right] \frac{m_\rpi^2 \Gamma(\rpi
\rightarrow \pho \pi) \Gamma(\rpi \rightarrow \r0 \pi)}{(m_\rpi^2-s)^2
+ m_\rpi^2 \Gamma_{tot}^2},
\ee
where $p_{cm}$ is the center of mass three-momentum of the incoming
particles, $s$ is the square of center of mass energy, $m_\rpi$ is the mass
of $\rpi$, the $\Gamma( A \rightarrow B C)$'s are momentum ($s$)
dependent widths to the incoming and outgoing channels,\footnote{Note
that C poses no relevant constraints on the decays of the charged
$R_\pi$ so both $R_{\pi^\pm} \rightarrow R^0 \pi^\pm$ and
$R_{\pi^\pm} \rightarrow \pho \pi^\pm$ are allowed even though the
$\r0$ and $\pho$ have opposite C eigenvalues.} and
$\Gamma_{tot}$ is the momentum dependent total width of the $\rpi$.
Thus, \eqr{eq:EQA} becomes $\avg{v
\sigma_{\r0 \pi}}=\avg{v \sigma_{\mbox{nonres}}} + \avg{v 
\sigma_{\mbox{res}}}$
where $\avg{v \sigma_{\mbox{nonres}}}$ is the formula given in
\eqr{eq:EQA} with $C$ set to a value of order 1.  Since we are concerned
with the maximum cross section resulting from the resonance, we are
focusing on the region of parameters for which the
non-resonant cross section is unimportant (i.e. $\avg{v
\sigma_{\mbox{nonres}}} \ll \avg{v \sigma_{\mbox{res}}}$).  

The kinematic momentum dependence of $\Gamma(A \rightarrow B C )$ can
be seen by expressing it in terms of a solid 
angle integral over the invariant amplitude squared $|{\cal M}|^2$:
\be
\Gamma( A \rightarrow B C ) = \left[p_{cm}(s)/(32 \pi^2 m_\rpi
\sqrt{s}) \right]
\int d\Omega |{\cal M}(A\rightarrow BC)|^2.
\ee
Here $p_{cm}(s) = 1/2 \sqrt{(s-(m_B + m_C)^2)(s-(m_B-m_C)^2)/s}$ is
the center of mass frame three-momentum of the decay products $B$ and
$C$.  Defining $4 \pi \xi_B \equiv \int d\Omega |{\cal M}(A\rightarrow
BC)|^2$ and assuming that $\Gamma_{tot} \approx \Gamma(\rpi \rightarrow
\r0 \pi)$, the three independently adjustable parameters for the
resonance are taken to be $m_\rpi, \xi_\r0$, and $\xi_\pho$.  

Since $R_\pi \rightarrow R^0 + \pi$ is a strong decay, we can
take its matrix element to be similar to the matrix element for
some known strongly decaying resonance whose decay has no angular
momentum barrier, for instance the $f_0(1370)$ whose total width is
300-500 MeV\re{PDG96}.  Thus we use $ \xi_\r0 \approx 16 \pi ~
\Gamma(f_0(1370))~ m_{f_0(1370)} = 7.4$ GeV$^2$.  

To determine $\xi_\pho$ we estimate the ratio of $\rpi \rightarrow
\pho \pi$ and $\rpi \rightarrow \r0 \pi$ matrix elements by keeping
track of the factors entering the short distance operator responsible
for $\r0 \rightarrow \pi \pi \pho$, namely $\pho \tilde{g} q \bar{q}$.
We use the $\r0$ mass, $M$, to set the scale.  This gives
\be
\label{xipho}
\xi_\pho/\xi_\r0 = A~ e^2 \frac{\alpha_s(M_S)M^4}{\alpha_s(M)M_S^4}
\approx A~ 4 \times 10^{-10} ~r^4 ~ \mu_8^4~ \mu_S^{-4}.
\label{eq:bbbb}
\ee

\begin{figure}
\hspace*{25pt} \epsfxsize=400pt \epsfbox{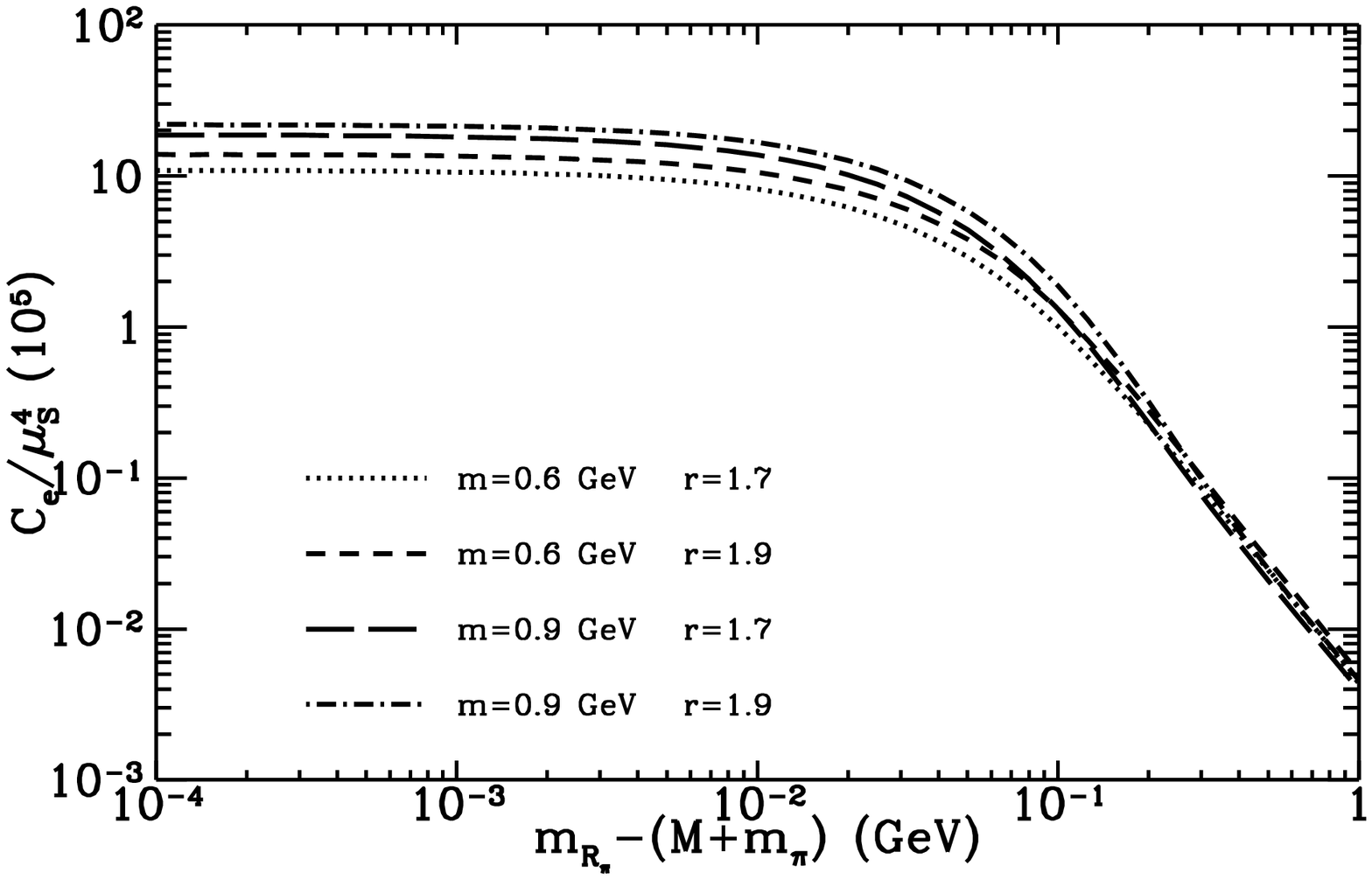}\\
\hspace*{1em} {\footnotesize{{\bf Fig.\ 5:} The effective resonance
enhancement factor $C_e/\mu_S^4$ is shown as a function of $r,m$, and
$m_\rpi$.}}
\end{figure}

We define $C_e$ to be the effective value of $C$ in \eqr{eq:EQA} that
would reproduce the $\Omega_\pho h^2$ calculated using the present
resonance model for a given set of resonance parameters, and taking
$A$ large enough to keep the $\r0$ in equilibrium abundance until
after photino-freeze out.  To calculate the thermal average $\avg{v
\sigma_{\mbox{res}}}$ of the resonant cross section, we use a
non-relativistic approximation which is within a factor of two or
better of the exact average.  It can be expressed in terms of the
one-dimensional integral
\be
\avg{v \sigma_{\mbox{res}}} = x^{3/2} \sqrt{\frac{8}{\pi}
\left(\frac{\mu_8}{0.175}+ \frac{1}{r}\right)} \int_0^{\infty} d\beta \beta
\sigma(s(\beta)) e^{-\beta x}
\label{eq:nonreltherm}
\ee 
where $s(\beta) \approx 0.64 \mu_8^2 (r+0.175/\mu_8)(r+0.175/\mu_8+2
\beta)$ GeV$^2$.  

$C_e$ is largest when the resonance is near the threshold
of the $\r0 \pi$ channel, because near the threshold $\Gamma(R_\pi
\rightarrow \r0 \pi)$ is phase space suppressed in comparison to
$\Gamma(R_\pi \rightarrow \pho \pi)$ and the peak value of the
Breit-Wigner cross section is proportional to $\sim \Gamma(R_\pi
\rightarrow \pho \pi)/\Gamma(R_\pi \rightarrow \r0 \pi)$.
However, because the width of the resonance vanishes as
$m_\rpi$ approaches threshold, the thermal average integral of the
Breit-Wigner cross section does not grow arbitrarily large.

In order to assess the plausibility of the original range used in
Ref.\ \re{farkolb}, we plot (Fig.\ 5) $C_e/\mu_S^4$ as a function of $r, m$,
and $m_\rpi$ with $A=1$ (in \eqr{eq:bbbb}) and $\mu_S=1/2$.  In all of
the $m$ and $r$ cases shown, $C_e/\mu_S^4 \ga 2 \times 10^5$ (or equivalently
$C_e \ga 10^4$) when $m_\rpi-M-m_\pi \la 70$ MeV.  Since the mass
splitting can easily be less than 70 MeV, we see that the $C$ range
used in Ref.\ \re{farkolb} would be inadequate, were self-annihilation
to be significantly larger than the non-resonant estimate adopted here.

\newpage

\frenchspacing
\def\prpts#1#2#3{Phys. Reports {\bf #1}, #2 (#3)}
\def\prl#1#2#3{Phys. Rev. Lett. {\bf #1}, #2 (#3)}
\def\prd#1#2#3{Phys. Rev. D {\bf #1}, #2 (#3)}
\def\plb#1#2#3{Phys. Lett. {\bf #1B}, #2 (#3)}
\def\npb#1#2#3{Nucl. Phys. {\bf B#1}, #2 (#3)}
\def\apj#1#2#3{Astrophys. J. {\bf #1}, #2 (#3)}
\def\apjl#1#2#3{Astrophys. J. Lett. {\bf #1}, #2 (#3)}
\begin{picture}(400,50)(0,0)
\put (50,0){\line(350,0){300}}
\end{picture}

\vspace{0.25in}

\def\labelenumi{[\theenumi]}

\begin{enumerate}

\item \label{bgmbm} R. Barbieri and L. Girardello and A. Masiero,
\plb{127}{429}{1983}; R. Barbieri and L. Maiani, \npb{243}{429}{1984}. 

\item \label{FM} G. R. Farrar and A. Masiero, Technical Report
RU-94-38 (hep-ph/9410401), Rutgers Univ., 1994.

\item \label{pierce_papa} D. Pierce and A. Papadopoulos, \npb{430}{278}{1994}.

\item \label{EXPTS} G. R. Farrar, \prd{51}{3904}{1995}.

\item\label{PHENO} G. R. Farrar,  Technical Report RU-95-17
(hep-ph/9504295), RU-95-25 (hep-ph/9508291), and RU-95-26
(hep-ph/9508292), Rutgers Univ., 1995.

\item \label{fPRL} G. R. Farrar, \prl{76}{4111}{1996}.

\item \label{aleph:lg}  The ALEPH Collaboration, Technical Report
CERN-PPE-97/002, CERN, 1997.

\item \label{fLaT} B. Gary, CTEQ Workshop, FNAL, Nov. 1996; G. R.
Farrar, Rencontres de la Valee d'Aoste, La Thuile, Feb. 1997 (preprint
in preparation).   

\item \label{farkolb} G. R. Farrar and E. W. Kolb, \prd{53}{2990}{1996}.

\item \label{SQUARKS} G. R. Farrar, \prl{76}{4115}{1996}.

\item \label{GG} P. Gondolo and G. Gelmini, \npb{360}{145}{1991}.

\item \label{KT} E. W. Kolb and M. S. Turner, {\em The Early
Universe}, (Addison-Wesley, Redwood City, Ca., 1990).

\item \label{agnello} Agnello et al., \plb{256}{349}{1991}. 

\item \label{G}  H. Goldberg, \prl{50}{1419}{1983}.

\item \label{LSP} J. Ellis, J. S. Hagelin, D. V. Nanopolous, K. Olive,
and M. Srednicki,  \npb{238}{453}{1984};
G. B. Gelmini, P. Gondolo, and E. Roulet,  \npb{351}{623}{1991};
K. Griest, \prd{38}{2357}{1988};
L. Rozkowski, \plb{262}{59}{1991}.

\item \label{INDIRECTA} J. Silk, K. Olive, and M. Srednicki,
\prl{53}{624}{1985};
T. K. Gaisser, G. Steigman, and S. Tilav, \prd{34}{2206}{1986};
J. S. Hagelin, K. W. Ng, and K. A. Olive,  \plb{180}{375}{1987};
M. Srednicki, K. A. Olive, and J. Silk, \npb{279}{804}{1987}.

\item\label{KKC}  G. L. Kane and I. Kani, \npb{277}{525}{1986};
B. A. Campbell {\em et al.}, \plb{173}{270}{1986}.

\item \label{goodwitt} M. W. Goodman and E. Witten, \prd{31}{3059}{1985}.

\item \label{PDG96} Particle Data Group, \prd{54}{1}{1996}.

\end{enumerate}

\end{document}